\documentclass[journal]{IEEEtran}
\IEEEoverridecommandlockouts
\usepackage{cite}
\usepackage{amsmath,amssymb,amsfonts}
\usepackage{algorithmic}
\usepackage{graphicx}
\usepackage{textcomp}
\usepackage{xcolor}
\usepackage{hyperref}
\def\BibTeX{{\rm B\kern-.05em{\sc i\kern-.025em b}\kern-.08em
    T\kern-.1667em\lower.7ex\hbox{E}\kern-.125emX}}
    
\usepackage{gensymb}
\usepackage{booktabs}
\usepackage{subfig}
\usepackage[load=addn]{siunitx}

\begin{document}

\title{Thin Film (High Temperature) Superconducting Radiofrequency Cavities for the Search of \\Axion Dark Matter\\

\thanks{This project has received funding from the European Union’s Horizon 2020 Research and Innovation programme under Grant Agreement No 730871 (ARIES-TNA). BD and JG acknowledge funding through the European Research Council under grant ERC-2018-StG-802836 (AxScale). We also acknowledge funding via the Spanish Agencia Estatal de Investigacion (AEI) and Fondo Europeo de Desarrollo Regional (FEDER) under project PID2019-108122GB-C33, and the grant FPI BES-2017-079787 (under project FPA-2016-76978-C3-2-P). Furthermore we acknowledge support from  SuMaTe RTI2018-095853-B-C21 from MICINN co-financed by the European Regional Development Fund, Center of Excellence award Severo Ochoa CEX2019-000917-S and CERN under Grant FCCGOV-CC-0208 (KE4947/ATS).\\
J.~Golm is with the European Organization for Nuclear Research (CERN), 1211 Geneva 23, Switzerland and the Friedrich - Schiller - University Jena, 07743 Jena, Germany (email: Jessica.Golm@cern.ch). \\
S.~Arguedas Cuendis, S. Calatroni, B.~D\"obrich, T. Koettig, J. Liberadzka-Porret, C.~Malbrunot, W. L. Millar,  C. Pereira Carlos, G. J. Rosaz, M. Siodlaczek, W. Wuensch are with the European Organization for Nuclear Research (CERN), 1211 Geneva 23, Switzerland. \\
X. Granados, J. Gutierrez, N. Lamas, T. Puig, G. Telles are with the Institut de Ciència de Materials de Barcelona, CSIC 08193 Bellaterra, Catalonia, Spain.\\
C.~Cogollos is with the Instituto de Ciencias del Cosmos, University of Barcelona, 08028 - Barcelona, Spain.\\
J.D.~Gallego is with Yebes Observatory, National Centre for Radioastronomy Technology and Geospace Applications, 19080 - Guadalajara, Spain.\\
 J.M. Garc\'ia Barcel\'o and P.~Navarro are with the Department of Information and Communications Technologies, Technical University of Cartagena, 30203 - Murcia, Spain.\\
 I.G.~Irastorza is with CAPA \& Departamento de F\'isica Te\'orica, University de Zaragoza, 50009 - Zaragoza, Spain.
}
}

\author{J.~Golm, S.~Arguedas Cuendis, S. Calatroni, C.~Cogollos, B.~D\"obrich, J.D.~Gallego, J.M. Garc\'ia Barcel\'o,  X. Granados, J. Gutierrez, I.G.~Irastorza, T. Koettig, N. Lamas, J. Liberadzka-Porret, C.~Malbrunot, W. L. Millar, P.~Navarro, C. Pereira Carlos, T. Puig, G. J. Rosaz, M. Siodlaczek, G. Telles and~ W. Wuensch}

\maketitle

\begin{abstract}
The axion is a hypothetical particle which is a candidate for cold dark matter. Haloscope experiments directly search for these particles in strong magnetic fields with RF cavities as detectors. The Relic Axion Detector Exploratory Setup (RADES) at CERN  in particular is searching for axion dark matter in a mass range above 30 $\mu$eV. The figure of merit of our detector depends linearly on the quality factor of the cavity and therefore we are researching the possibility of coating our cavities with different superconducting materials to increase the quality factor. Since the experiment operates in strong magnetic fields of 11 T and more, superconductors with high critical magnetic fields are necessary. Suitable materials for this application are for example REBa$_2$Cu$_3$O$_{7-x}$, Nb$_3$Sn or NbN. \\
We designed a microwave cavity which resonates at around 9~GHz,  with a geometry optimized to facilitate superconducting coating and designed to fit in the bore of available high-field accelerator magnets at CERN. Several prototypes of this cavity were coated with different superconducting materials, employing different coating techniques. These prototypes were characterized in strong magnetic fields at 4.2 K.

\end{abstract}

\begin{IEEEkeywords}
Superconducting resonators, SRF superconducting radio frequency cavities, Quality factor, 2G HTS Conductors,axion
\end{IEEEkeywords}

\section{Introduction}
 \IEEEPARstart{T}{he} Relic Axion Detector Exploratory Setup (RADES) is an axion haloscope experiment searching for dark matter axions in a strong magnetic fields with high quality factor cavities. It differs from most haloscopes in the fact that thus far it employs dipole magnets and not solenoids. {Many experiments successfully use copper cavities in strong magnetic fields to set limits to the axion coupling at low mass ranges.} Reference \cite{Irastorza:2018dyq} { and \cite{Semertzidis:2021rxs}} and references therein provide a recent review of experimental axion searches and the haloscope technique. {The past years superconducting cavities were explored by experiments like QUAX \cite{QUAX1},\cite{QUAX2} and CAPP \cite{CAPP,Ahn:2021fgb} in order to reach a higher sensitivity.} An axion with mass  $m_A$  converts into a  photon  due to the inverse Primakoff effect. If the converted photon's energy matches the resonance frequency of the cavity, the output power is augmented depending on the axion's coupling strength to photons.
 For a given axion-photon coupling the figure of merit $F$ of the experiment is
 \begin{equation}
\label{eq:SNR}
    \text{$F$} \sim g_{a\gamma}^2 m_{A}^2B^4V^2 T_{sys}^{-2} G^4 Q ,
\end{equation}
 where $g_{a\gamma}$ is the axion coupling to two photons, $B$ the external magnetic field (assumed constant over the cavity volume), $V$ is the cavity volume, $T_{sys}$ is the detection noise temperature, and $G$ is the geometric form factor of the cavity mode.
 
 The figure of merit of the experiment increases by the power of four with the strength of the magnetic field. Therefore we aim at magnets with fields as high as possible. For the current run we had a 2-m long \SI{11}{\tesla} dipole magnet in single coil configuration available, for details see \cite{Magnet}. This sets the requirements for the coatings: we needed a type II superconductor with a critical magnetic field $B_{c2}$ well above \SI{11}{\tesla} at \SI{4.2}{\kelvin}. {The materials should also possess a RF surface resistance {R$_s$} lower than copper at our operating conditions. Experimental results and theoretical predictions have been described in literature, see for example \cite{Barcelona}, \cite{Silva} and references therein for Nb$_3$Sn or high temperature superconductors like REBa$_2$Cu$_3$O$_{7-x}$ (RE = Y, Gd, Eu) (REBCO).} Both these materials were applied to our cavities. One cavity was sputter-coated with Nb$_3$Sn and REBCO tapes were applied to the second cavity, where the hastelloy substrate was stripped off, such that the REBCO layer is exposed to the RF fields. 

\section{Cavity design}
\begin{figure}[htb]
\centering

\subfloat[]{
	\label{subfig:geometry}
	\includegraphics[width=0.37\textwidth]{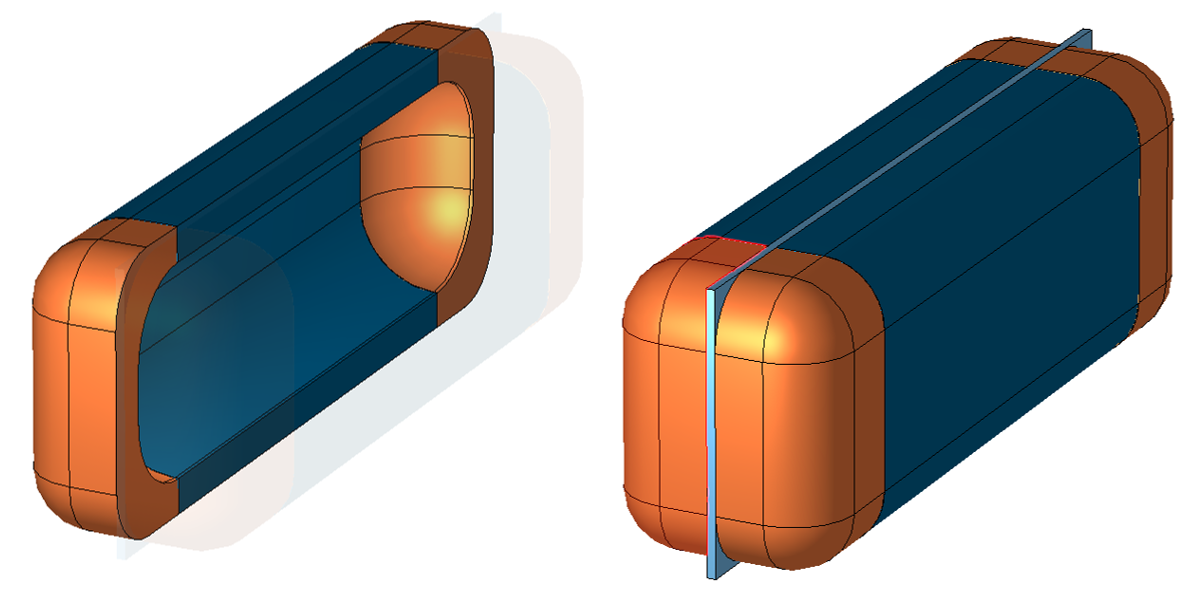}} 

\subfloat[]{
	\label{subfig:surfacecurrents}
	\includegraphics[width=0.36\textwidth]{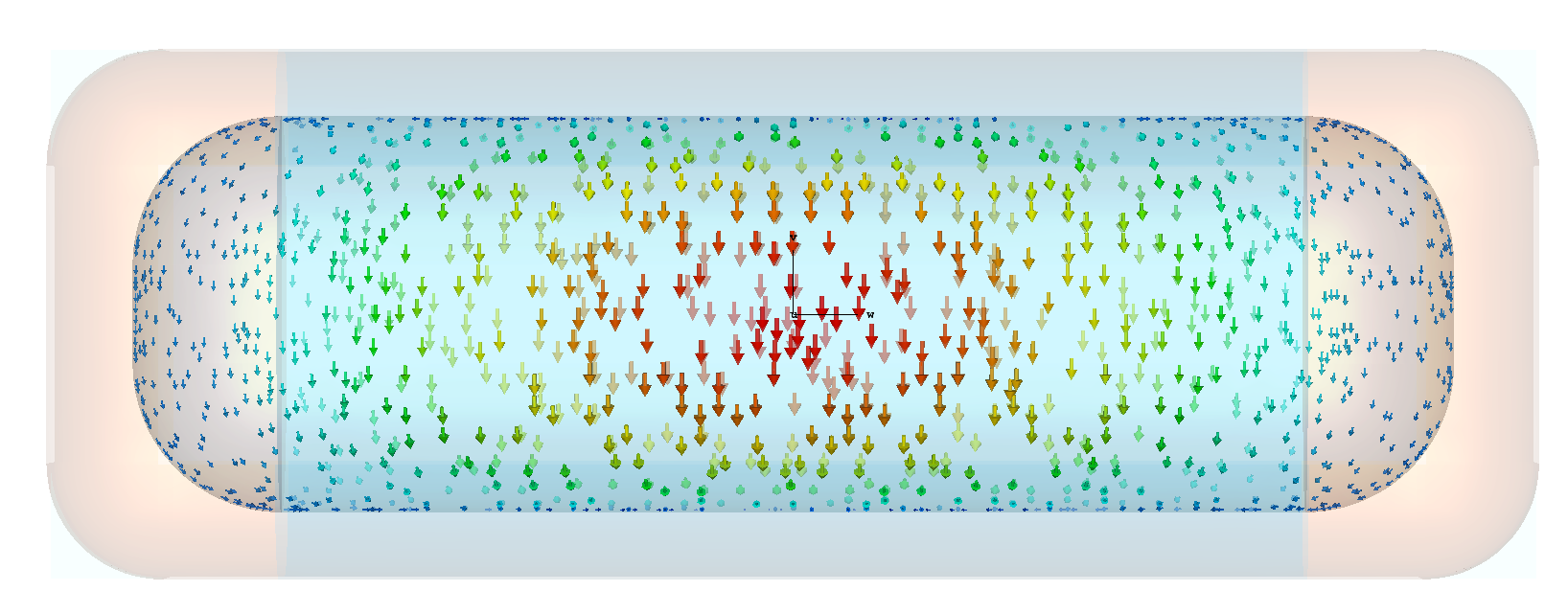}} 
	
\caption{Cavity design modeled in CST {Studio Suite®} showing (a) cavity halves design and (b) surface currents of the axion mode. }
\label{fig:Cavity_geometry}

\end{figure}

 The previous RADES design consists of rectangular sub-cavities joined by irises and was specially designed for the axion search in long dipole magnets \cite{Melcon:2018dba}. To facilitate the coating of the cavities the design had to be adapted: the irises have been removed and the corners rounded. The cavity is designed to have a large volume (given the frequency), high geometric and quality factor, resonate at about \SI{9}{\giga\hertz} while fitting in the common bores of dipole magnets that were at the time available for RADES experiments. The optimization of the cavity geometry was done in CST and the final design is shown in Fig. \ref{fig:Cavity_geometry}. The cavity is cut along the electrical field lines shown in Fig. \ref{fig:Cavity_geometry} {(a)}. The cut at this position allows us to tune the cavity at a later stage. {No RF currents flow across the gap,} and the cavity halves can be separated up to \SI{2}{\milli\metre} without degrading sizeably the quality factor \cite{InPrep}. In Fig. \ref{fig:Cavity_geometry} {(b)} the direction of the surface currents is illustrated. The superconducting tape was attached to the walls in the flow direction of the currents. In this way the currents do not have to cross the tape boundaries which may result in an increase of the surface resistance across this boundary.\\
 For Nb$_3$Sn cavity halves the entire inner surface was coated with the superconductor, but the layer in the rounded corners may not be optimally coated because of the angle to the sputtering source. However, as shown in Fig. \ref{fig:Cavity_geometry} {(b)} there is also not much current flowing around this corners. So even if the coating does not reach its optimum in this corners this will only lead to a small degradation in Q$_0$.\\
 For the cavity coated with an HTS tape it was even decided to omit these rounded corners as shown in orange in Fig.~\ref{fig:Cavity_geometry}~{(a)}. Instead the area was chosen to be copper coated. The simulations showed that for HTS with a surface resistance five times better than copper, the degradation in Q$_0$ resulting from using copper for the corners is only about \SI{8}{\percent}. \\ 
 {The superconducting layer is not expected to shield significantly the magnetic field, due to its low thickness, as demonstrated in \cite{EUCAS2021}.}
\section{Coatings}
\subsection{Cu coating}
\begin{figure}[htb]
    \centering
\centering

\subfloat[]{
	\label{subfig:Cumat}
	\includegraphics[width=0.34\textwidth]{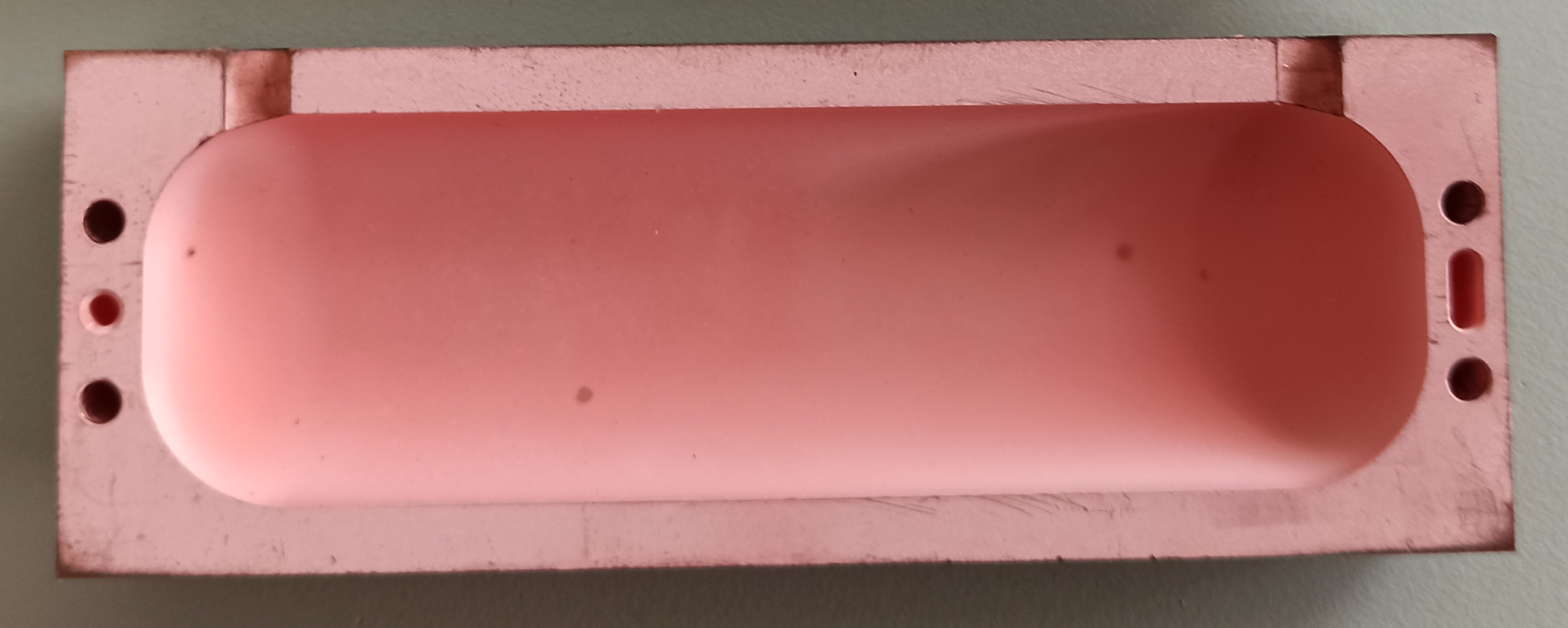}} 

\subfloat[]{
	\label{subfig:Cushiny}
	\includegraphics[width=0.34\textwidth]{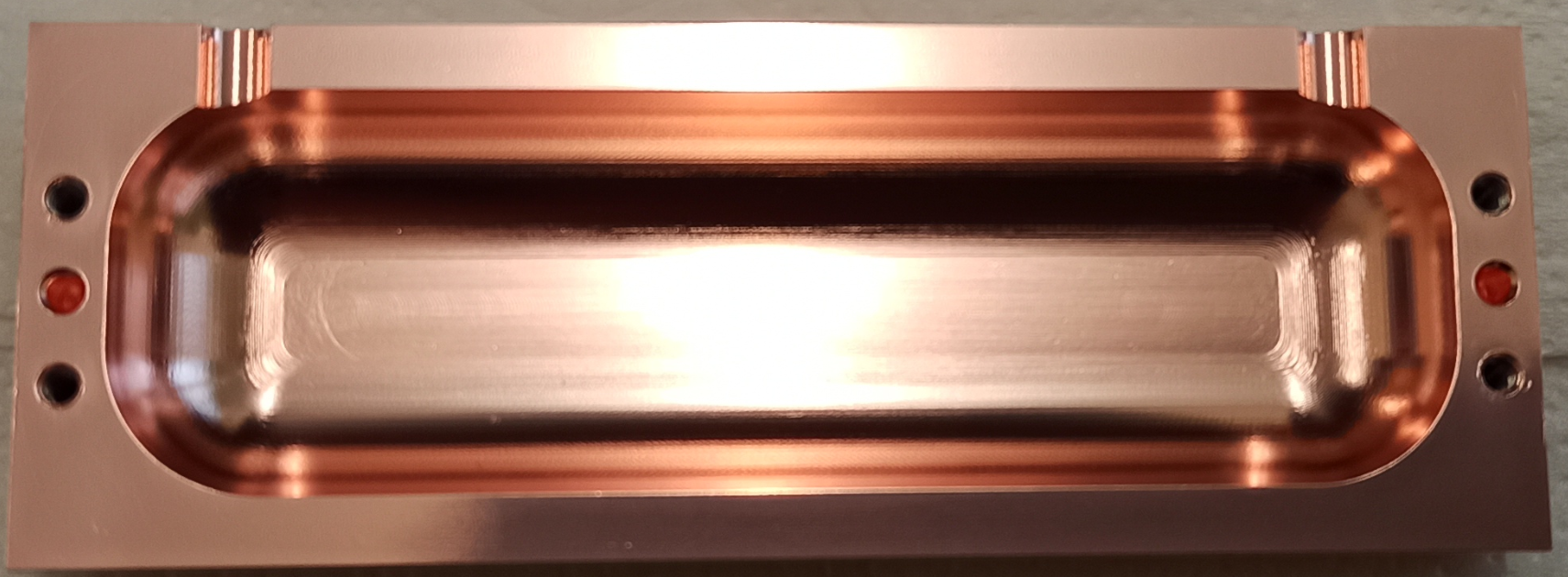}} 
	
 \caption{Photographs of a RADES cavity half coated with matt (a) and shiny (b) copper.  }
 \label{fig:CuCavities}

\end{figure}

A reference cavity {of the same geometry as the superconducting cavities} was coated with copper of two different versions - `shiny' and `matt'. The matt copper coating was applied by pulsed galvanic plating and {has a high purity (Residual-resistance ratio (RRR) of 680$\pm$100 \cite{Amador_2021}) but the surface roughness is very high and it appears matt. } The shiny coating was applied by DC galvanic plating with an organic brightener. {The surface roughness is much smaller for this coating method but RRR is only around 38$\pm$2 \cite{Amador_2021}.} Fig. \ref{fig:CuCavities} shows photographs of a RADES cavity half with the different copper coatings applied. Since we are operating at frequencies around \SI{9}{\giga\hertz}, the quality factor of the cavities at low temperatures is limited by the anomalous skin effect. Due to this effect the surface roughness {matters more than the purity of the material}. For that reason{,} we measured a much higher quality factor for the shiny copper cavity at \SI{4.2}{\kelvin} {of} about 40{,}000 (R$_s$ = {6.6 m$\Omega$}) than for the matt one {of} about 30{,}000 (R$_s$ = {8.8 m$\Omega$}). {The s}hiny copper plating was retained for our reference cavity, and was applied also to the HTS cavity before soldering the tapes.

\subsection{Nb$_3$Sn {coating}}
\begin{figure}[h!]
\centering

\subfloat[]{
	\label{subfig:correct}
	\includegraphics[width=0.28\textwidth]{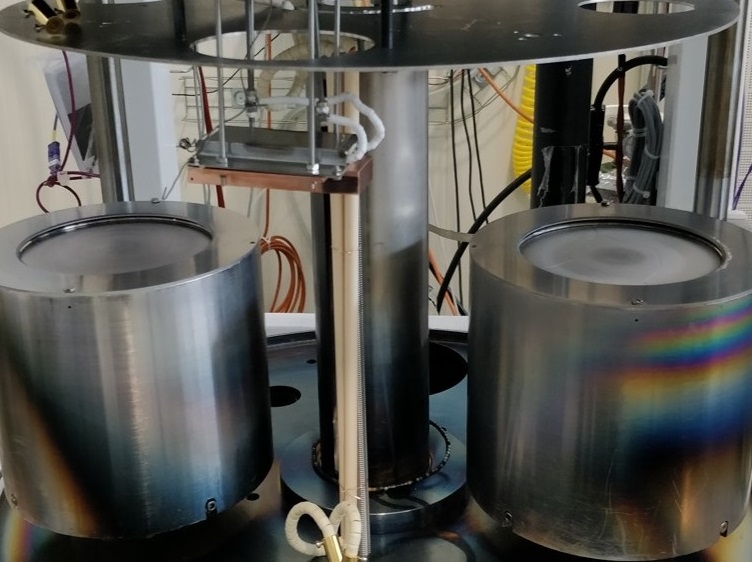}} 

\subfloat[]{
	\label{subfig:notwhitelight}
	\includegraphics[width=0.28\textwidth]{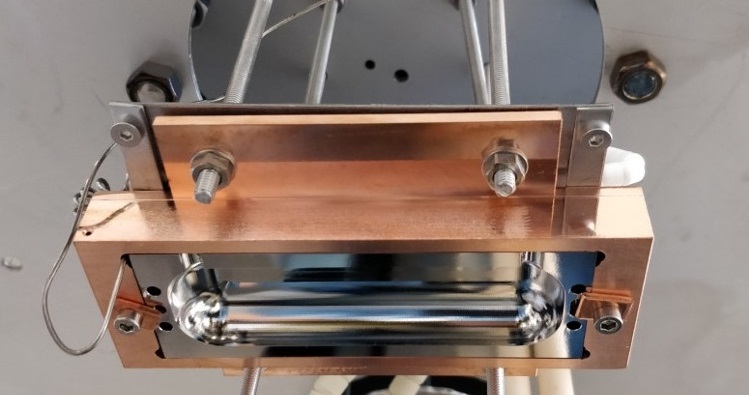}} 
	
\caption{Photographs of the (a) coating setup and (b) a cavity half in coating position.}
\label{fig:Nb3Sn coating setup}

\end{figure}

The cavity halves were coated separately in two subsequent runs maintaining the same parameters. For each cavity half, a Ta intermediate layer was grown onto the substrates (electropolished stainless steel 316LN) to avoid diffusion of the substrate within the Nb$_3$Sn layer, coated immediately after. This two-step process was made possible by the dual-magnetron configuration of our coating set up, one with a Ta and one with a Nb–Sn stoichiometric target. A custom Joule heater was built in-house to keep the sub-cavities at \SI{750}{\celsius} throughout the entire coating and annealing processes. Photographs of (a) the coating setup with both magnetrons and the sample holder in the centre, with the homemade heater attached on top, and (b) a close up of a cavity half in coating position (previous to coating) are shown in Fig. \ref{fig:Nb3Sn coating setup}. 

The Ta layer was deposited by High-Power Impulse Magnetron Sputtering with a potential reversal for ion acceleration (HiPIMS + Positive Pulse) for \SI{40}{\minute} at \SI{1e-3}{\milli\bar}, resulting in $\approx$ \SI{0.9}{\micro\metre} thick layers. After this step, each cavity half was left at the same temperature and pressure conditions (\SI{750}{\celsius}, \SI{1e-3}{\milli\bar} of Kr) for a \SI{45}{\minute} anneal to promote the formation of Ta-$\alpha$ phase. The Nb$_3$Sn layer, $\approx$ \SI{2}{\micro\metre} thick, was sputtered resorting to Direct Current Magnetron Sputtering (DCMS) for \SI{75}{\minute} at a Kr pressure of \SI{7e-4}{\milli\bar}. The final step for each of the cavity halves was a longer second anneal lasting around \SI{24}{h} in order to promote the formation of the Nb$_3$Sn superconducting phase and cure the potential defects \cite{Ilyina_2019}.

The coating parameters (duration $t$, pressure $p$, temperature $T$ and power $Pw$)  used for the deposition of each layer are summarized in Table \ref{tab:Nb3Sn coating parametres}. The HiPIMS Main Pulse (duration $t$ and frequency $\nu$) and applied Positive Pulse (duration $t$ and delay $\Delta$t) input values used for the Ta layer are also presented.

\begin{table}[t!]
\centering
\caption{Coating parameters for the Ta and Nb$_3$Sn layers.}
\label{tab:Nb3Sn coating parametres}
\begin{tabular}{@{}p{0.02\textwidth}cccccccc@{}}
\toprule
 & \textbf{t} & \textbf{p} & \textbf{Pw} & \textbf{T} & \multicolumn{2}{c}{\textbf{Main pulse}} & \multicolumn{2}{c}{\textbf{Positive Pulse}} \\
 & (min) & (mbar) & (W) & \textbf{($\degree$C)} & t & $\nu$ & t & $\Delta$t \\
 &  &  &  &  & ($\mu$s) & (kHz) & ($\mu$s) & ($\mu$s) \\ \midrule
\textbf{Ta} & 40 & 1$\times$10$^{-3}$ & 350 & 750 & 50  & 1 & 200 & 4 \\
\textbf{Nb$_3$Sn} & 75 & 7$\times$10$^{-4}$ & 350 & 750 & - & - & - & - \\ \bottomrule
\end{tabular}
\end{table}

\subsection{REBCO tape}
\begin{figure}[htb]
\centering

\subfloat[]{
	\label{subfig:HTSschematics}
	\includegraphics[width=0.36\textwidth]{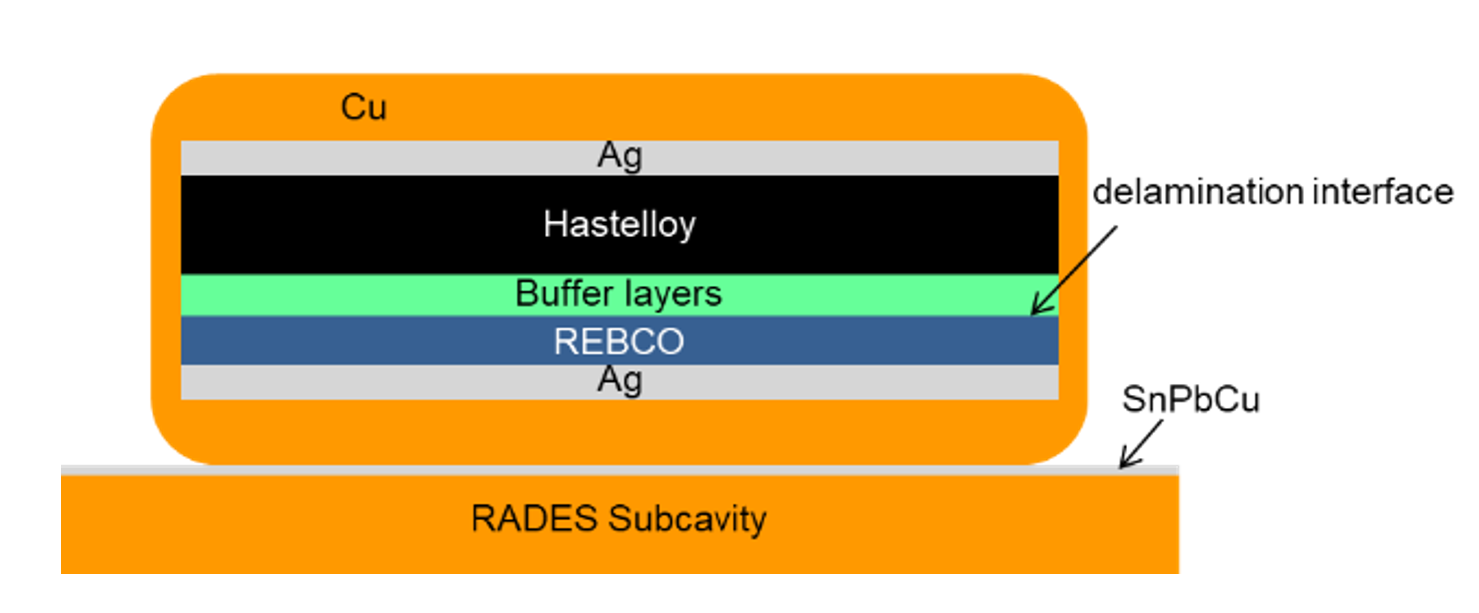}} 

\subfloat[]{
	\label{subfig:HTSpic}
	\includegraphics[width=0.35\textwidth]{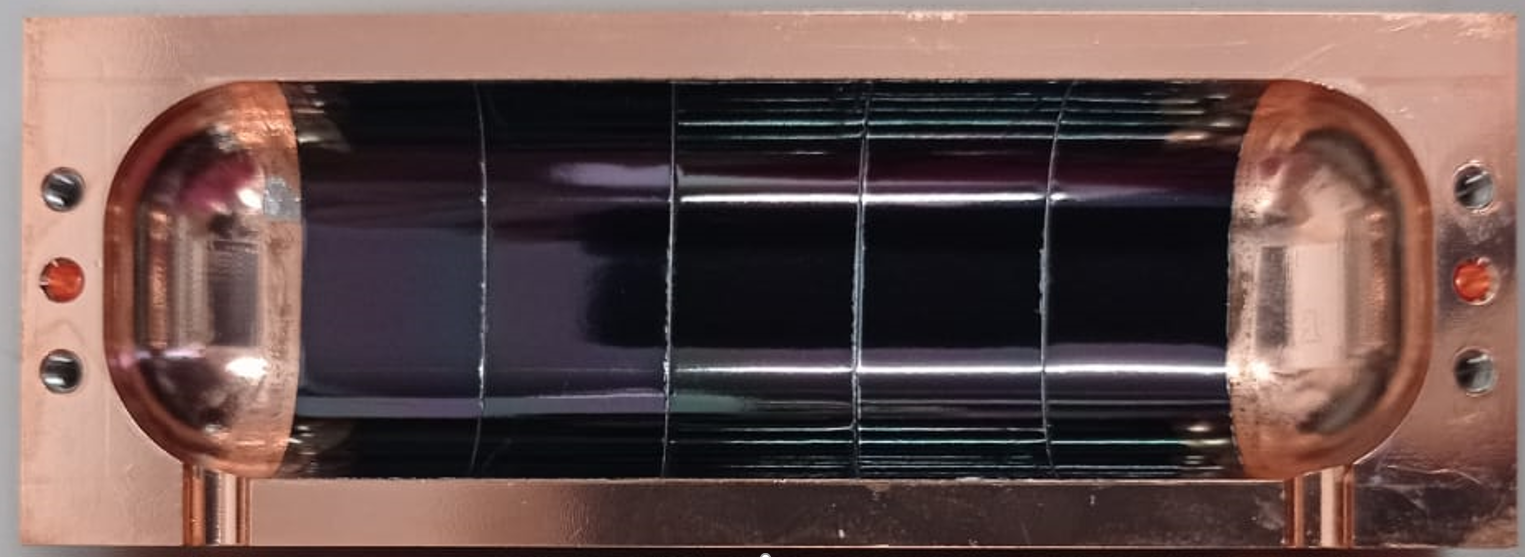}} 
	
\caption{Schematics of the {(a)} coated conductor segment's position on the RADES cavity and picture of    one of the {(b)} RADES halves after the coating is completed. }
\label{fig:HTScavity}

\end{figure}

Each cavity half was coated with 5 segments of 12 mm-wide coated conductor provided by the company THEVA. The segments were placed in a way that the REBCO layer was oriented towards the surface of the cavity half (see Fig. \ref{fig:HTScavity} {(a)}). To homogeneously heat up the system, we used specially designed flexible heaters that could adapt to the curved geometry of the cavity halves. The coated conductors were soldered onto the surface of the cavity halves using Sn$_{60}$Pb$_{38}$Cu$_2$ due to its low-temperature melting point. {A standard procedure has been applied for such low temperature solders that enables short heating times of less than 5 minutes at T $<$~260 ºC.  During the development of the soldering technique, samples of 12 x 50 mm$^2$ where characterized by means of a scanning Hall probe microscope (SHPM) before and after soldering. The soldering technique guarantees no degradation to the REBCO material within the experimental error of our SHPM technique, which we estimate to be under 10 \%.} During the soldering process, the heaters and the tapes were kept into position by applying some {small mechanical} pressure onto the system. Once the system cooled down to room temperature, the tapes were peeled off by pulling out the hastelloy substrate. In that way {the buffer layer delaminates from the REBCO surface}, (see Fig. \ref{fig:HTScavity} {(b)}).

\section{Experimental set-up}
\begin{figure}[htb]
    \centering
	\includegraphics[width=0.46\textwidth]{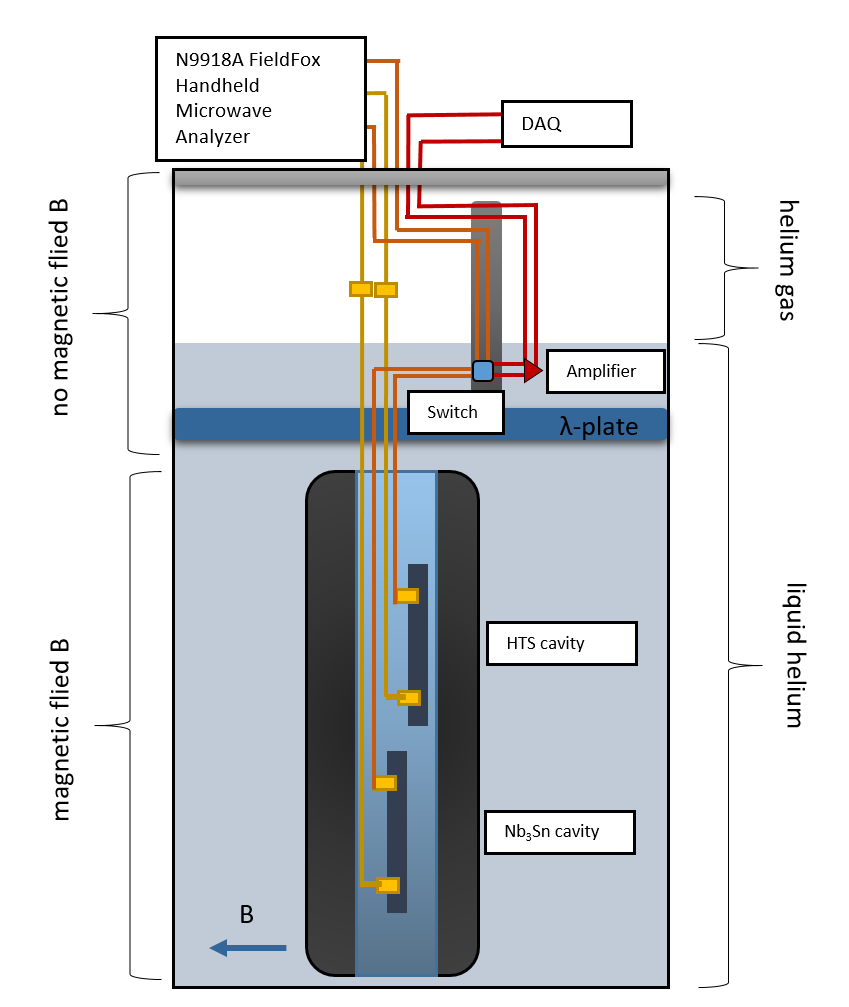} 
 \caption{Schematics of the experimental set-up for the quality factor measurements and axion data taking in a \SI{11}{\tesla} magnetic field at CERN.  }
 \label{fig:SM18}

\end{figure}

Before testing the new coatings in magnetic field, the quality factor was measured in the CERN Central Cryogenic Laboratory in liquid helium at \SI{4.2}{\kelvin}. For the Nb$_3$Sn cavity we could measure a $Q_0$ of about 700{,}000 (R$_s$ = {0.4 m$\Omega$}) and for the HTS tape about 80{,}000 (R$_s$ = {3.3 m$\Omega$}), both values being much higher than the $Q_0$ of 40{,}000 from our copper reference cavity. \\
Afterwards the cavity was installed in a \SI{11}{\tesla} dipole short model at CERN and the characterization of the cavities in high magnetic field was done in parallel to an axion search. One cavity port is weakly coupled (and terminated for axion search) and for the other port we aim for a critical coupling. 
Since the power of a signal that may be generated by photons converted from axions is very small, a low noise amplifier is connected to the critically coupled port. After amplification the signal is detected by a custom-made data acquisition system from TTI \cite{TTI}. \\
In order to be able to measure the quality factor in this configuration a switch was necessary to bypass the low noise amplifier. The signal from the critical coupled port goes from the switch to one port of a vector network analyser while the weakly coupled port goes directly to the device. \\
The magnet bore was filled with liquid helium which has a dielectric constant of 1.049343  at \SI{4.2}{\kelvin} \cite{DielLHe}. The insertion into a dielectric de-tunes our cavity and results in a resonance frequency of about \SI{8.8}{\giga\hertz} during data-taking. Fig. \ref{fig:SM18} shows the experimental set-up. Two cavities were installed at the same time in the magnet bore of \SI{54}{\milli\metre} diameter. The bottom cavity was the Nb$_3$Sn coated cavity and the top one the HTS-tape cavity. A Hall sensor was attached onto the bottom cavity in order to align the cavity with the magnetic field.

\section{Results}
\begin{figure}[htb] 
	\begin{minipage}{\columnwidth}
		\includegraphics[width=0.95\textwidth]{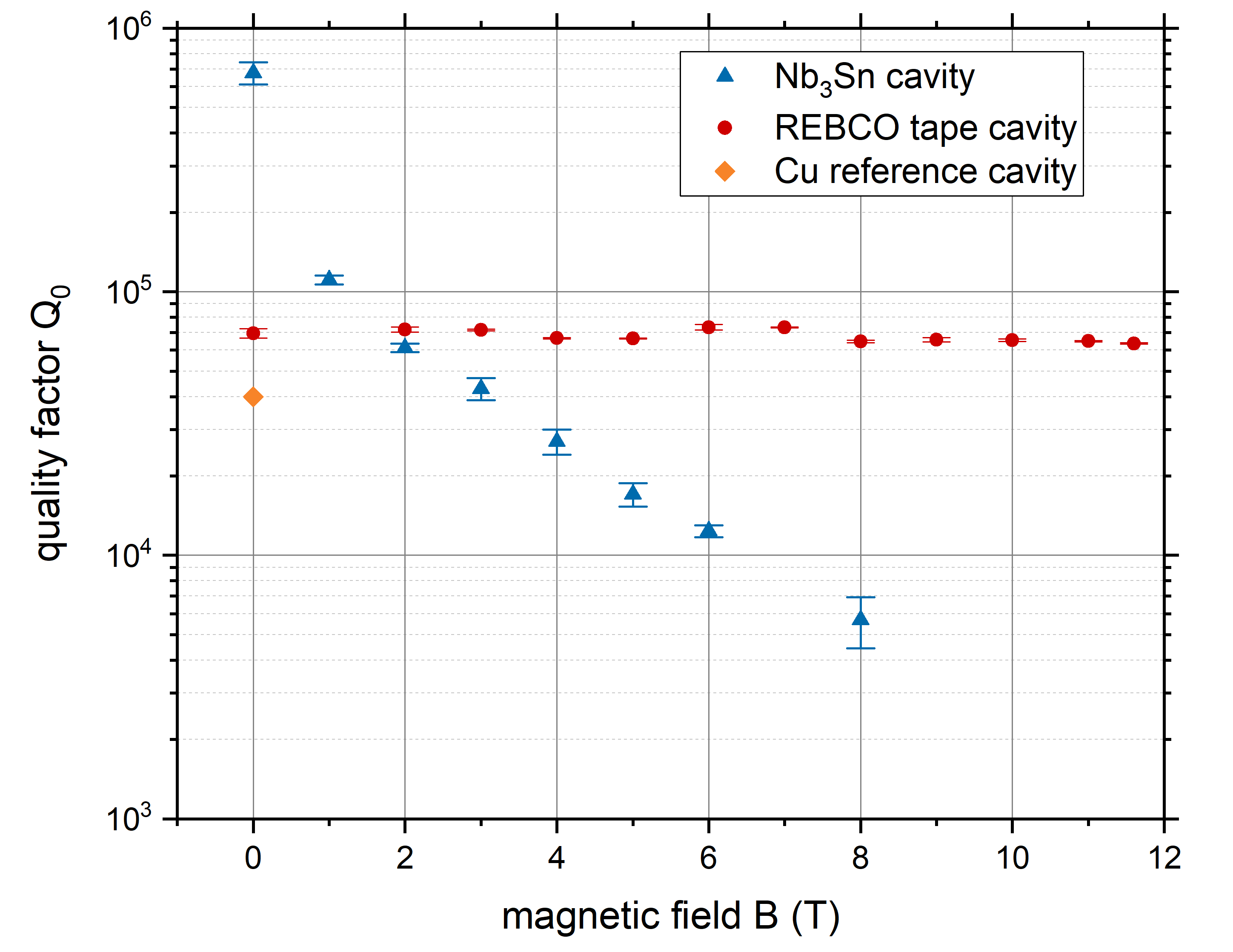}
	\end{minipage}
	\caption{Results of quality factor measurements with the cavity immersed in liquid helium.{ The resonance frequency of the cavities is f = (8.8 \SI{ \pm  0.1}) {\giga\hertz}} at 4.5 K.}
	\label{fig:2}
\end{figure}

The quality factor for each cavity was calculated from the S-parameter measurements with a vector network analyser using the 3dB method 
to determine $Q_{l}$: 

 \begin{equation}
\label{eq:Ql}
    \text Q_{ l }=\frac{ f_{ 0} }{ \Delta f_{ 3dB } } ,
\end{equation}

where $f_{ 0 }$ is the frequency of the maximum amplitude and $\Delta f_{ 3dB }$ the bandwidth at -~3~dB. The coupling of the strongly coupled port was determined using the reflection parameters.  
During the measurement in liquid helium we observed a fast frequency drift which interfered with the Q measurement and suspect pressure fluctuations to be responsible for this shift. For each measurement we recorded a frequency range of \SI{2}{\mega\hertz} measuring 10{,}001 points in this range. The frequency sweep took about 6 seconds and the drift in the frequency within this time is reflected by the error bars of the Q values in Fig. \ref{fig:2}. 
The magnetic field was ramped up in \SI{1}{\tesla} steps with a speed of \SI{10}{\ampere/\second} (about { 1~kA/T}). Afterwards the field was kept constant for 10 minutes and the quality factor of both cavities was measured. The results for the $Q_0$ of both cavities are shown in Fig. \ref{fig:2}. The quality factor of the HTS tape cavity remained almost constant between 60{,}000 (R$_s$ = {4.4 m$\Omega$})and 80{,}000 (R$_s$ = {3.3 m$\Omega$})up to \SI{11.6}{\tesla}, while Nb$_3$Sn decreased considerably and performed worse than our copper reference cavity above \SI{3}{\tesla}. Investigations about this behaviour are ongoing. 
The HTS cavity outperformed the copper cavity by 50\% in quality factor and increased the sensitivity of our axion data-taking {since the figure of merit scales linear with Q, see equation \eqref{eq:SNR}. } These physics results will be the object of a separate publication. {From results presented in \cite{Barcelona} we expected an improvement of the quality factor by a factor of 5. After the initial characterization of the quality factor of the cavity our studies have shown that the decrease in quality factor has its origin in the 9 mm curvature radius of the cavity. Currently, we are working on the coating of a second haloscope in which we are implementing an upgraded procedure and we expect to obtain a significant improvement in the quality factor}.\\

\section*{Acknowledgment}

We are indebted to the CERN technical teams, particularly  Gerard Willering, Franco Julio Mangiarotti, Jerome Feuvrier, Marta Bajko, {Patrick Viret, Guillaume Pichon, Arnaud Devred,} Stephan Russenschuck and Andrzej Siemko, as well as the CERN cryolab team. We also thank the RADES team and Kristof Schmieden for helpful discussions {and Giuseppe Ruoso for the loan of equipment used in the measurement.}

\end{document}